# TEMPERATURE DEPENDENT STRUCTURE AND DYNAMICS OF ISOPROPANOL-WATER LIQUID MIXTURES AT LOW ALCOHOL CONTENT

*Szilvia Pothoczki*[1], *László Pusztai*[1,2] *and Imre Bakó*[3]

[1] Wigner Research Centre for Physics, Hungarian Academy of Sciences, H-1121 Budapest, Konkoly Thege út 29-33., Hungary

[2] International Research Organization for Advanced Science and Technology (IROAST), Kumamoto University, 2-39-1 Kurokami, Chuo-ku, Kumamoto, 860-8555., Japan

[3] Research Centre for Natural Sciences, Hungarian Academy of Sciences, H-1117 Budapest, Magyar tudósok körútja 2., Hungary

**Abstract**

Series of molecular dynamics simulations for 2-propanol-water mixtures, as a function of temperature (between freezing and room temperature) and composition ($x_{ip}$= 0, 0.5, 0.1 and 0.2) have been performed for temperatures reported in the only available experimental structure study. It is shown that when the all-atom OPLS-AA interatomic potentials for the alcohol are combined with the TIP4P/2005 water model then near-quantitative agreement with measured X-ray data, in the reciprocal space, can be achieved. Such an agreement justifies detailed investigations of structural, energetic and dynamic properties on the basis of the simulation trajectories. Here we focus on characteristics related to hydrogen bonds (HB): cluster-, and in particular, ring formation, energy distributions and lifetimes of HB-s have been scrutinized for the entire system, as well as for the water and isopropanol subsystems. It is demonstrated that, similarly to ethanol-water mixtures, the occurrence of 5-membered hydrogen bonded rings is significant, particularly at higher alcohol concentrations. Concerning HB energetics, an intriguing double maximum appears on the alcohol-alcohol HB energy distribution function. HB lifetimes have been found significantly longer in the mixtures than they are in the pure liquids.

*E-mail: pothoczki.szilvia@wigner.mta.hu; Phone: +36 1 392 1469

*E-mail: bako.imre@ttk.mta.hu; Phone: +36 1 382 6981

## 1. INTRODUCTION

Aqueous solutions of alcohols, the latter showing both hydrophobic and hydrophilic characters, have provided an excellent testing ground, from many aspects, for scientific research for many decades[1-24]. Additionally, these solutions are of basic importance in numerous fields, ranging from fundamental science to industrial applications.

Recently, the temperature dependent structure of methanol/water[25] and ethanol/water[26] liquid mixtures, as well as the microscopic dynamics[27] in ethanol/water solutions has been investigated extensively in water-rich mixtures, using molecular dynamics (MD) simulations. The basis for detailed discussions of structure and dynamics has been provided by very good agreements with measured[15] X-ray diffraction data.

The temperature-dependent experimental work of Takamuku[15] extends to aqueous solutions of yet another alcohol, isopropanol (a.k.a. 2-propanol, or propan-2-ol), again, at low alcohol concentrations (between molar ratios, $x_{ip}$, of 0.05 and 0.2). Since findings of our previous investigations[25-27] have proven to be rather thought-provoking, we have decided to follow on with a detailed T-dependent MD study on isopropanol/water mixtures.

The first explanation to the perturbation of a hydrophobic probe (originally alcohols) induced to the water structure was proposed Frank and Evans in 1945[1], and was based mainly on thermodynamic properties of these solution. Since that time, the proposed 'iceberg' formation of water molecules around hydrophobic solutes has provoked a significant debate in the scientific community. The main conclusion of this hypothesis is that water in the solvation shell of a hydrophobic species (a molecule, or part of a molecule) has a somewhat higher degree of hydrogen bonding than bulk water does. However, there is still no consensus concerning this hypothesis, and about details of the atomistic picture that cause the non-ideal behaviour in terms several of their macroscopic properties at low alcohol concentrations. 2-propanol is the simplest example of a secondary alcohol, where the alcohol carbon atom is attached to two other carbon atoms and in this sense, the OH-group of this molecule has a compact hydrophobic environment.

It is known[28-35] that various physicochemical quantities like enthalpy of mixing, dielectric properties, diffusion constant, excess molar volume have minimum values for 2-propanol/water mixtures at low alcohol concentration, as well as for other short chain length alcohols[3,9,11,13,14,28,31,33,35]. The rapid change in the dynamical properties of water and alcohol (self-diffusion coefficients, rotational correlation time) in the water-rich region (up to about 0.15-0.2 mole fraction)[12,32] indicates that there is a significant structural change in the solution, in comparison with pure liquid water. From earlier studies[6,10] it is possible to conclude that with increasing the size of the nonpolar head group of alcohols, more pronounced effects on different thermodynamic quantities and dynamical properties can be observed. Multiple studies indicate that alcohol molecules (n-propanol[17], tert-butanol[36], 1,1,1-3,3,3 hexfluoro-2-propanol (HFIP)[18]) aggregate in their aqueous solutions, according to the small angle X-ray (SAXS) and neutron (SANS) scattering data in the water rich region[19-21]. On the other hand, the measurable correlation length is significantly smaller in the 2-propanol-water system than for the 1-propanol (size is the almost the same as 2-propanol) or tert-butanol-water case[17,22,36]. Very similar difference was observed in terms of the dynamical properties between mixtures of 1-propanol and 2-propanol with water, by Sato et al[28]. This is a serious indication that besides the size of the hydrophobic group, an additional important factor, the possible formation of various hydrogen bonded aggregates, is present that may govern the structural and dynamical change in these solutions.

There are several works that explore the structural and dynamical properties of isopropanol/water liquid mixtures using molecular dynamics simulation[12,23,30]. These simulations showed that recently employed force fields provide acceptable agreement over a wide composition range between calculated and measurable physico-chemical properties. There has been no direct comparison reported for the structural properties (total radial distribution functions and/or total structure factors), not even in the most recent simulation study on these mixtures[30]. A possible reason for this hiatus is that most of the newly developed potential functions are of the 'united atom' type (see, e.g. Ref. 30) that do not consider hydrogen atoms individually.

It is widely accepted that the perturbation of the hydrogen-bond (HB) network is one of the reasons behind the anomalous properties[1,6,8]. The structure of these systems at the molecular level may be described using a local structural parameter like average H-bonding number or hydrogen bond distribution. On the other hand, we can characterise these systems as a complex networks, which, in turn, may be described by their topological properties. Recently, some of the present authors have described the topology of hydrogen bonded aggregations in water and pure liquid formamide[37-39]. It could be demonstrated that although both alcohol and water molecules form hydrogen bonds readily, significant differences are present in terms of the H-bonded environment of the two species in water-methanol and water-ethanol mixtures.

Here we consider 2-propanol-water mixtures, with 2-propanol contents of 5, 10 and 20 mol%, at temperatures between ambient and the freezing point of the actual mixture. We validated our simulation procedure by comparing measured[15] and calculated total structure factors. One of the aims of the present study was to obtain new insights into the hydrogen bonded network of 2-propanol-water mixtures. We analyse quite a few characteristics as a function of decreasing temperature, such as size distributions of cyclic entities or the size of H-bonded aggregates. The other goal of the present work was to describe and localise changes of the interaction energy between the constituent molecules. To this end, we provide a more detailed picture of the energetics of the interactions around water and 2-propanol molecules, in 2 and 3 dimensions. Additionally, we also study how the hydrogen bond lifetime changes in different hydrogen bonded environments, also as a function of temperature.

## 2. COMPUTATIONAL METHODS

### 2.1 Molecular Dynamics Simulations

Molecular Dynamics simulations were performed by the Gromacs software[40] (version 5.1.1). 2-propanol molecules were modelled using the all-atom optimized potentials for liquid simulations (OPLS-AA)[41] force field. Bond lengths were kept fixed by the LINCS algorithm[42] in 2-propanol molecules. Parameters such as atom types and charges can be found in Table S1. Calculations with two different water models, SPC/E[43] and TIP4P/2005[44], have been conducted for every composition-temperature pair. Water molecules were held together by the SETTLE[45] algorithm. Additional parameters like temperatures, box lengths, numbers of 2-propanol and water molecules in each system, number densities (and bulk densities), are summarized in Table S2. The Newtonian equations of motions were integrated via the leapfrog algorithm, using a time step of 2 fs. The particle-mesh Ewald algorithm was used for handling the long-range electrostatic forces and potentials.[46-47] The cut-off radius for non-bonded interactions was set to 1.1 nm.

The following simulation sequence was applied: first NPT systems (at each concentration) was heated up to 340 K, using a Nose-Hoover[48,49] thermostat with a time constant of $\tau_T=1.0$ and a Parrinello-Rahman[50] barostat with a time constant of $\tau_p=4.0$, over 5 ns to avoid the aggregation of 2-propanol molecules. After that, a 5 ns NVT equilibration run with a Berendsen[51] thermostat ($\tau_T=0.5$) was applied. This was the starting point for further simulations. By this sequence, it is possible to exploit that the Berendsen method is a fast, first-order approach to equilibrium, whereas the Nose-Hoover thermostat with Parinello-Rahman barostat provides canonical ensembles with correct fluctuation properties. Furthermore, in NVT simulations it is a good practice to perform the equilibration using the Berendsen thermostat with a small value of $\tau_T$, that should be increased later to obtain a stable trajectory in equilibrium.[52]

Accordingly, for every composition the following four steps were performed to reach the next, lower, temperature: 1. *NPT_short run* (2ns, Berendsen thermostat with $\tau_T=0.1$, Berendsen barostat with $\tau_p=0.1$), 2. *NPT_long run* (10ns, Nose-Hoover thermostat with $\tau_T=1.0$, Parrinello-Rahman barostat with $\tau_p=4.0$), 3. *NVT_short run* (1ns, Berendsen thermostat with $\tau_T=0.1$), 4. *NVT_long run* (5ns, Berendsen thermostat with $\tau_T=0.5$). All results were calculated from the NVT_long runs.

### 2.2 Analysis tools

For calculating partial radial distribution functions the *g_rdf* software was used, which can be found in the GROMACS software package. Total scattering structure factors were calculated from partial radial distribution functions by an in-house code.

Mean square deviations (MSD) were determined by the help of the *g_msd* software that is also included in GROMACS simulation package.

Analyses concerning hydrogen bonding, including energetic aspects, were performed using our in-house computer code, described in detail in Ref 37.

## 3. RESULTS

### 3.1 Structure

#### 3.1.1 Validation of molecular dynamics results: comparison with experimental total scattering structure factors

Measured X-ray diffraction data[15] are compared to molecular dynamics model total scattering structure factors (TSSF-s) in Fig. 1. Agreement with experiment is rather good for both water potential models applied: for the TIP4P/2005 model[44], the match is nearly quantitative. Contrary to what was found for ethanol-water mixtures in our earlier work[26], here this water potential works significantly better and therefore, in what follows, all results are shown for these calculations only.

Fig.1 shows TSSF-s only for one composition, $x_{ip}=0.2$ (i.e., 20 mol % isopropanol); graphs for the other compositions, as well as corresponding numerical data describing deviations between experiment and simulation, are provided in the Supporting Information.

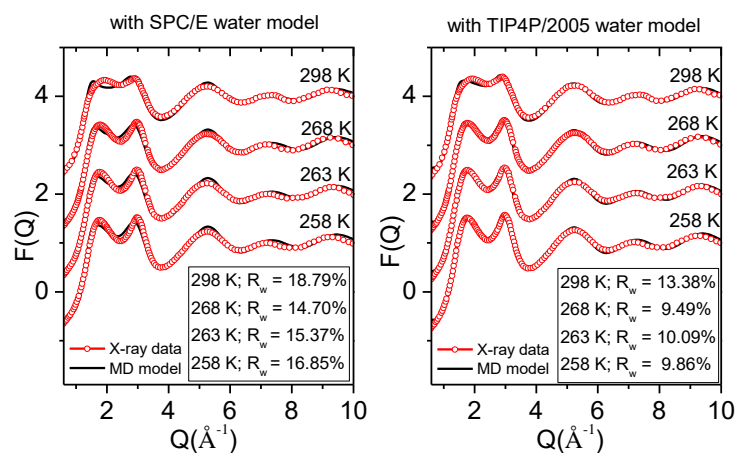

Figure 1 Temperature dependent experimental (red symbols, from Ref. 15) and computed (black solid line, present work; left panel: SPC/E, right panel: TIP4P/2005 water models) total scattering structure factors. Note that very nearly quantitative agreement between measurement and model when using the TIP4P/2005 water model (right panel).

Partial radial distribution functions can be found in the Supporting Information, so that the applicability of the criteria used for defining hydrogen bonds (see below) may be checked there.

3.1.2 Hydrogen bond statistics

The analysis of hydrogen bond statistics can give us information about the average local structure of a liquid in terms of a distribution of the number of molecules in positions forming hydrogen bonds with the central one. In the present study, the energetic definition[53] of the H-bond was applied as follows: two molecules were considered hydrogen bonded to each other if they were found at a distance $r(O\cdots H) < 2.5$ Å, and the interaction energy is smaller than -3 kcal/mol (ca. -12 kJ/mol). This definition has less arbitrariness than the pure geometrical definition, as we showed in an earlier publication[53], where all analysis using both the energetic and the geometric definition ($r(O\cdots H) < 2.5$ Å, and H-O…O angle between 30°) were carried out. It was found that the results arising from the two different definitions were in good agreement with each other, thus the main conclusions did not depend on the applied definitions.

First, we calculated the average number of H-bonds between all the molecules, and also separately for the contributions from water-water, water-2-propanol and 2-propanol-2-propanol pairs. Results are shown in Fig. 2 for all the three concentrations ($x_{ip}$=0.05, 0.1 and 0.2) and at all studied temperatures. In the molecule-molecule, water-water and water-2 propanol cases the average number of H-bonds increases as temperature decreases. Pairs of water molecules form the most H-bonds between each other. The average number of H-bonds is not shown in the figure for 2-propanol-2-propanol molecules, because less than one H-bond for such pairs are found and their number does not change significantly with temperature.

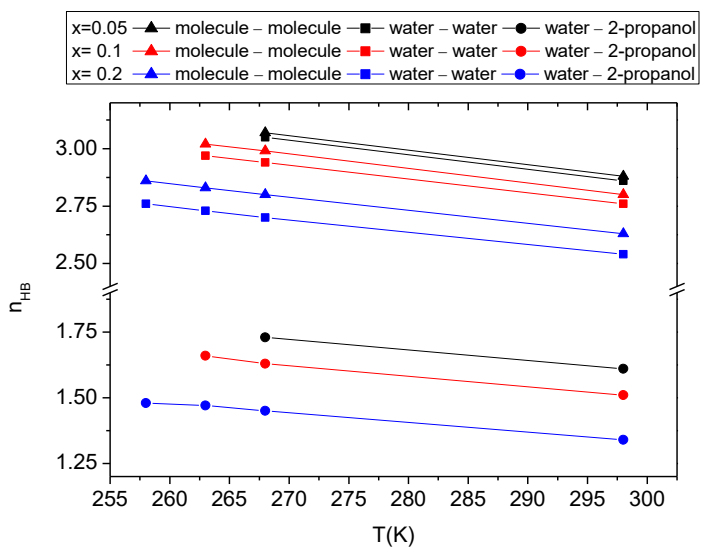

Figure 2 Average number of H-bonds.

Molecules can be classified based on the number of hydrogen-bonds they take part in as H-acceptors ($n_A$=0, 1, 2 for 2-propanol and $n_A$=0, 1, 2, 3 for water molecules) and H-donors ($n_D$=0, 1 for 2-propanol and $n_D$=0, 1 or 2 for water). Molecules may thus be tagged as ($n_A$,$n_D$), e.g. (0,0) – no bond, up to (3,2) – fully bonded. Calculated data are presented in Table 1. For all concentrations the fraction of 4 H-bonded water molecules (2,2) increases and the 2 H-bonded water molecules (1,1) decreases as temperature decreases. There is a well-defined asymmetry between (1,2) and (2,1) type water molecules in terms of their populations, and this difference is the same for every temperature in all concentrations. Concerning 2-propanol, the fraction of (1,2) increases while (0,1) decreases with decreasing temperature. In this regard our conclusion coincides what was found for ethanol-water systems: both 2-propanol and water molecules tend to form the maximum number of H-bonds allowed for them as a result of cooling.

| $x_{ip}$ | T(K) | water | | | | 2-propanol | | | |
|---|---|---|---|---|---|---|---|---|---|
| | | 1D:1A | 1D:2A | 2D:1A | 2D:2A | 0D:1A | 0D:2A | 1D:1A | 1D:2A |
| 0.2 | 298 | 0.182 | 0.149 | 0.278 | 0.249 | 0.317 | 0.121 | 0.334 | 0.126 |
| | 268 | 0.14 | 0.147 | 0.29 | 0.325 | 0.292 | 0.138 | 0.331 | 0.16 |
| | 263 | 0.133 | 0.145 | 0.29 | 0.339 | 0.28 | 0.138 | 0.34 | 0.168 |
| | 258 | 0.125 | 0.144 | 0.288 | 0.355 | 0.283 | 0.136 | 0.34 | 0.169 |
| 0.1 | 298 | 0.174 | 0.167 | 0.256 | 0.267 | 0.329 | 0.165 | 0.266 | 0.147 |
| | 268 | 0.13 | 0.166 | 0.258 | 0.353 | 0.288 | 0.191 | 0.267 | 0.185 |
| | 263 | 0.123 | 0.163 | 0.257 | 0.37 | 0.281 | 0.199 | 0.261 | 0.192 |
| 0.05 | 298 | 0.171 | 0.175 | 0.241 | 0.271 | 0.325 | 0.199 | 0.235 | 0.151 |
| | 268 | 0.129 | 0.176 | 0.24 | 0.357 | 0.285 | 0.288 | 0.222 | 0.195 |

Table 1 Fractions of water and 2-propanol Molecules as H-Acceptors and as H-Donors in the H-Bonds Identified as a Function of Temperature and Concentration.

3.1.3 Ring size distributions

Molecules participate in a given cyclic entity if there is a minimum length path consisting of a series of hydrogen bonds ($n_r$) through which one can get back to the original molecules. To estimate the ratio the cycle and the open chain in the system ring search algorithms developed by Chihaia et al.[55] were used. Here, the primitive rings of oxygen atom nodes and hydrogen bond edges were sought. This method has been already used for investigating the topology of H-bonded clusters in pure water, and in water-methanol, water-ethanol and water-formamide mixtures[25,26,38]. Ring size distributions (not normalized) for the three mixtures (left panel: $x_{ip}$ = 0.05; middle panel: $x_{ip}$ = 0.1; right panel: $x_{ip}$ = 0.2) are presented in Fig. 3, as a function of temperature.

In the case of pure water, the ring-size distribution has a well-defined maximum around 6 and the number of cyclic entities is significantly increasing as the temperature is decreasing[37]. A similar tendency was found in water-methanol mixtures as a function of temperature[15], and for ethanol water-mixtures below the alcohol concentration of $x_e=0.1$[26]. Here, in isopropanol-water mixtures, a clear dominance (but only by a narrow margin) of 6-membered rings is observed only at low temperature, and only when the alcohol molar ratio is below 0.1 (cf. Fig. 3).

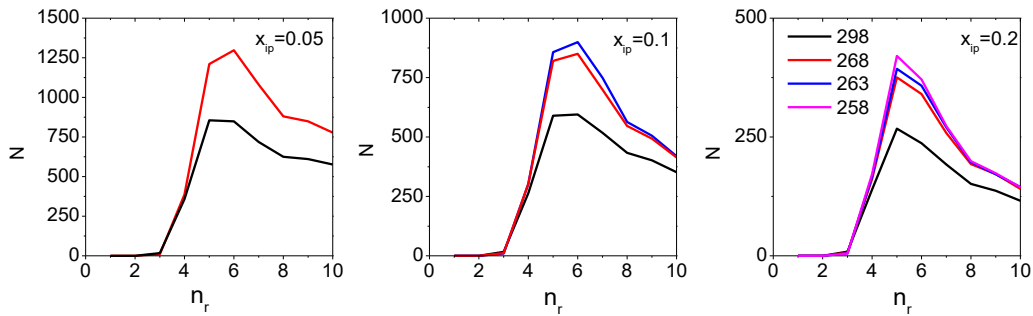

Figure 3 Non-normalized ring size ($n_r$) distributions with mean-square deviations (MSDs) for the three mixtures (left panel: $x_{ip}$ = 0.05; middle panel: $x_{ip}$ = 0.1; right panel: $x_{ip}$ = 0.2) as a function of temperature. H-bonds have been identified on the basis of the energetic criterion (*y* axes: "*N*": number of rings/configuration).

On the other hand, we already observed for ethanol-water mixtures that above a certain concentration ($x_e$=0.1), the most probable cyclic entities contain 5 molecules[26]. Also, in the present case the visible maximum of the ring size distributions shifts from 6- to 5-membered cycles as concentration increases. At the highest 2-propanol concentration ($x_{ip}$=0.2), 5-fold rings take the dominance already at room temperature, and their ratio systematically increases with lowering the temperature. A possible reason for this variation may be, as noted already for ethanol-water mixtures[26], that as the ratio of isopropanol molecules grows, the number of alcohol molecules in the ring structures, that need larger volumes if included, also grows. Simply the size of the hydrophobic isopropyl group seems to be sufficient to force the H-bonded ring to close sooner (i.e., with fewer molecules in the ring), by excluding molecules from the ring being formed.

A possibility of an interesting comparison shows up here: it is instructive to compare the 'relative importance' of cyclic structures in water, and ethanol-water and isopropanol-water liquid mixtures. For this purpose, we have normalized the number of rings by the number of molecules in the given system for each molar ratio in the three kinds of systems. Clearly, the largest ratio of ring structures is found for pure (TIP4P/2005) water at each temperature. The ratio is the smallest in isopropanol-water mixtures. This way, the influence of the size of the alkyl-group in the alcohol molecules on the ability of forming cyclic entities in alcohol-water mixtures could be demonstrated straightforwardly. The data are presented in Table 2.

|       | water | ethanol-water | | isopropanol-water | |
|-------|-------|------|------|------|------|
| x     |       | 0.1  | 0.2  | 0.1  | 0.2  |
| 298 K | 1.015 | 1.007 | 0.744 | 0.411 | 0.320 |
| 268 K | 1.177 | 1.152 | 0.899 | 0.570 | 0.430 |
| 263 K | 1.214 |       |       | 0.590 | 0.450 |
| 258 K | 1.282 | 1.255 | 0.939 |       | 0.475 |
| 253 K | 1.392 | 1.353 | 0.966 |       |       |

Table 2 Ratio of the number of rings and the total number of molecules in pure water, and in ethanol-water and isopropanol-water mixtures.

In water-methanol and water-ethanol mixtures we previously detected that more alcohol molecules appear in non-cyclic associations, while more water molecules are connected to rings, than it would follow from the composition. In order to provide a clear distinction between the behavior isopropanol and water in these mixtures, we calculated the average number of 2-propanol ($n_{ip}$) molecules incorporated in cyclic structures of certain sizes. If the H-bonding character of water and 2-propanol molecules was identical at this level then this number should be approximately equal to $n_r \cdot x_e$, where $n_r$ is the size of the ring. Results obtained are presented in Fig 4. The calculated values at every concentration and temperature are significantly smaller than $n_r \cdot x_e$. (Fig 4a and 4b) The deviation from the ideal behavior is less pronounced in the case of smaller rings. The deviation from ideal behavior, as measured by the ratio of the calculated and the ideal value (Fig 4c and 4d), has a well-defined minimum at all concentration and temperature around $n_r=7$. At low temperature the deviation from ideality is more pronounced at room temperature for larger ($n_r > 7$) rings.

To sum up, as far as cyclic structures are concerned, 2-propanol/water mixtures are in a closer relation with ethanol/water, than with methanol/water systems. This observation may be explained by the smaller size ratio of isopropyl vs. ethyl (ca. 3/2), than ethyl vs. methyl (ca. 2/1) alkyl groups.

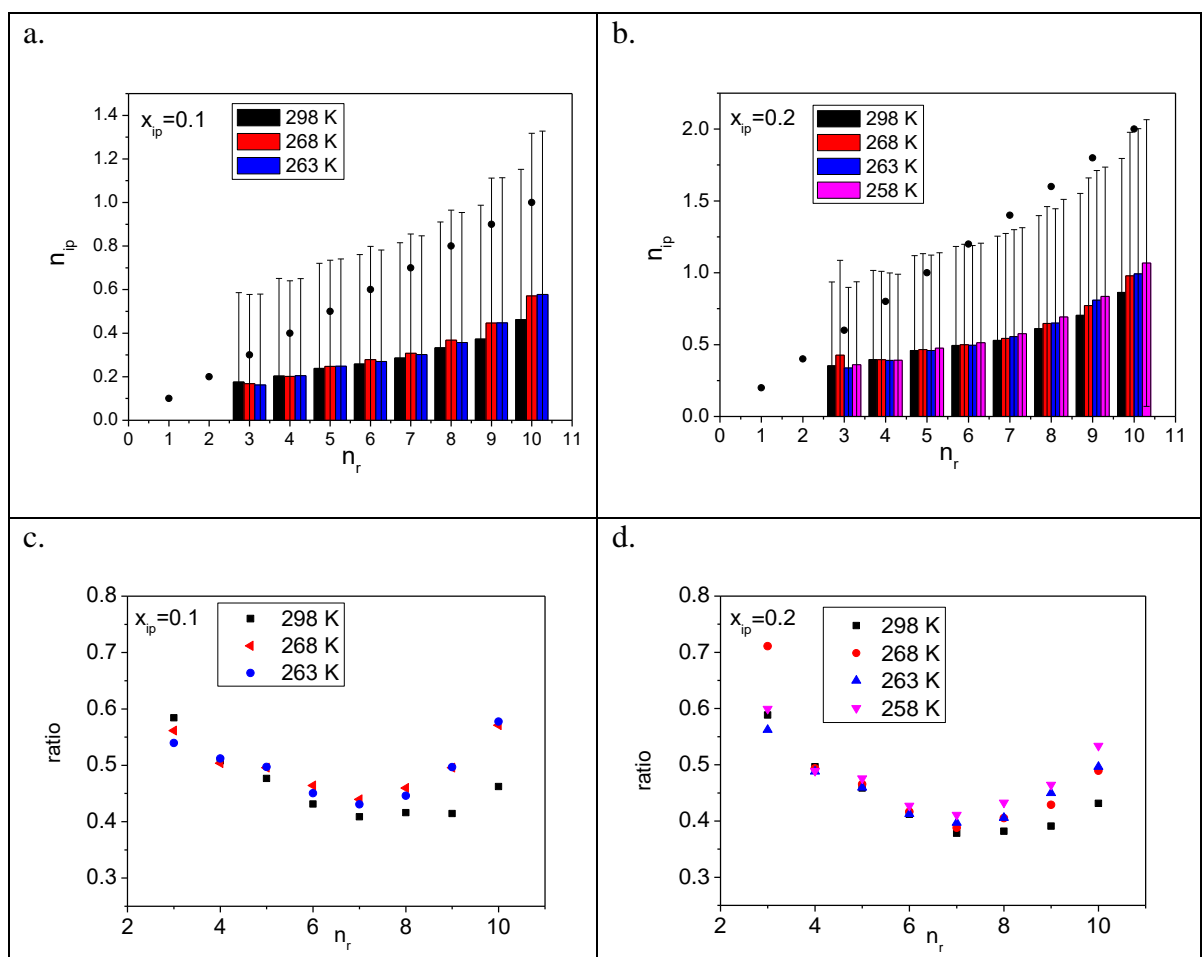

Figure 4 Average number of isopropanol molecules incorporated in cyclic structures of certain sizes, as a function of temperature and composition for $x_{ip}=0.1$ (a) and for $x_{ip}=0.2$ (b). (Black dots denote values that would follow directly from the given alcohol concentration).

The ratios of the calculated and the ideal values are also provided for $x_{ip}=0.1$ (c) and for $x_{ip}=0.2$ (d).

### 3.1.4 On the formation of H-bonded clusters

In order to reveal the possible existence of large clusters formed by water and 2-propanol molecules through H-bonds, we calculated the cluster size distributions for the water and 2-propanol subsystems, along with the case when both components were counted as cluster formers. In these analyses two molecules are regarded as belonging to the same cluster if they are connected by a chain of hydrogen bonds between molecules of the type of interest. Percolation can be assigned by comparing the calculated cluster size distribution functions of the present systems with those obtained for random percolation on a 3D cubic lattice (see Refs. 56-58), $P(n_c)= n_c^{-2.2}$. In percolating systems, the cluster size distribution ($P(n_c)$) exceeds this predicted function at large cluster size ($n_c$) values.

Figure 5 presents the hydrogen-bonded cluster size distributions in 2-propanol-water mixtures at various concentrations for 'molecule-molecule', water-water and 2-propanol-2-propanol subsystems. It is clear that the 'molecule-molecule' (entire system), as well as the water subsystems percolate through the simulation box. On the other hand, we can find only short chain-like structures (consisting of less than 10 molecules) for pure 2-propanol assemblies.

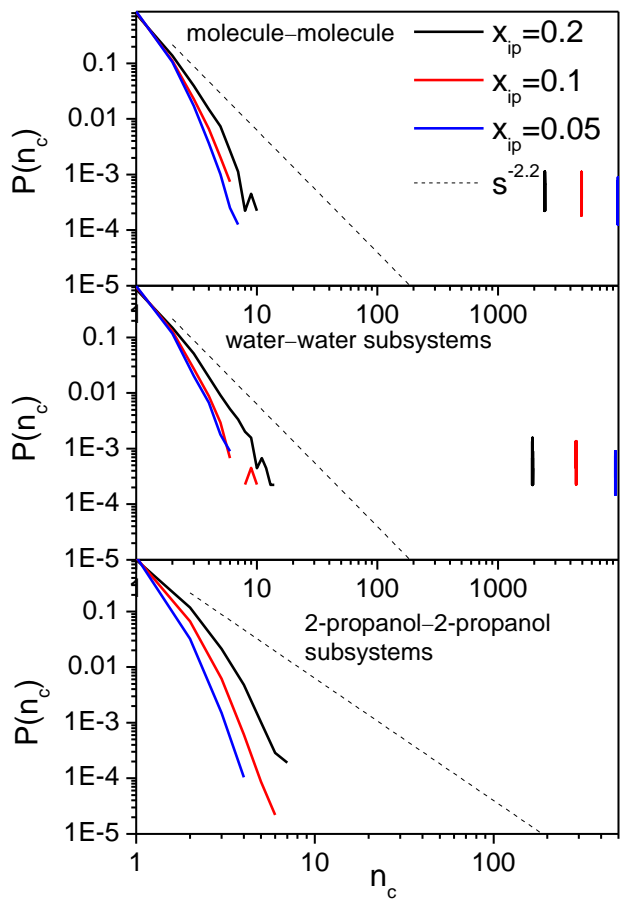

Figure 5 Cluster size distributions as calculated for the entire system ('molecule-molecule', top panel), as well as for the water-water (mid panel) and 2-propanol-2-propanol (bottom panel) subsystems.

3.2 Hydrogen bond energetics

It is well established that for a more complete understanding of the properties of aqueous mixtures at the molecular level, it is beneficial to make use of various statistical tools concerning not only the structural, but also the dynamical and the energetic aspects. Following this idea, here we first analyze the strength of intermolecular connections between water and 2-propanol molecules via the pair energy distributions (Fig. 6) in their mixtures, as a function of composition and temperature.

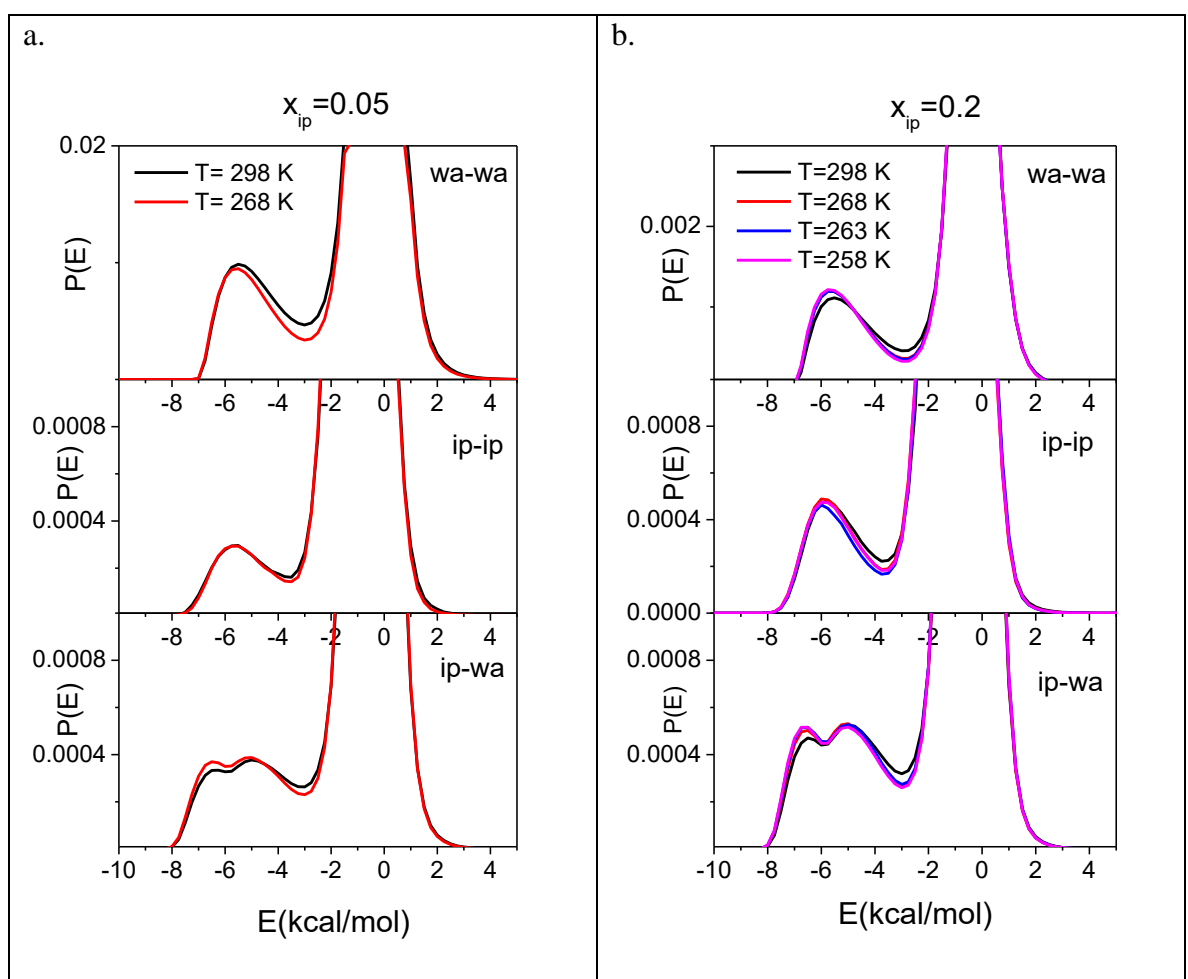

Figure 6 Pair energy distributions in 2-propanol-water mixtures at $x_{ip}$=0.05 (a) and 0.2 (b) at different temperatures.

Pair energy distributions in H-bonded liquids usually have a characteristic shape (see, e.g., Refs. 26 and 38), with (1) a spike near 0.0 kcal/mol that represents the interaction between distant molecules in the bulk, and (2) a low energy band for hydrogen bonded neighbors (following the first, well-defined minimum). The distributions of pair energies for water-water ('wa-wa') and 2-propanol-2-propanol ('ip-ip') interactions all exhibit a peak at negative E values, at around –5.5 kcal/mol for water-water and –5.75 kcal/mol for 2-propanol-2-propanol connections, respectively. The positions of these maxima are shifted to more negative values with decreasing temperature. This statement is also valid for the well-defined minima in the cases of the two compositions shown ($x_{ip}$=0.05 and $x_{ip}$=0.2) that can be found at -3.0 kcal/mol for water-water and at -3.5 kcal/mol for 2-propanol-2-propanol pairs.

The 2-propanol-water pair energy distributions show two small maxima around -5.2 kcal/mol and -6.5 kcal/mol. These two peaks can be assigned to H-bond donor and acceptor interactions, which differ in their strengths, between 2-propanol and water molecules. It is worth pointing out that this double peak was not found in ethanol-water mixtures (cf. Ref. 26): this is a striking difference between ethanol-water and 2-propanol-water mixtures. At all temperatures and for all concentrations the minimum after these maxima can be found at -3.0 kcal/mol.

In order to establish how the changes in terms of the H-bond interactions described above relate to the distances between molecules, we have calculated the pair interaction energy as a function of the O-O distances for all of the three molecular pair combinations: water-water, 2-propanol-2-propanol and 2-propanol-water. Although we calculated these functions for all concentrations and at each simulated temperature value (c.f Table 1), in Figures 7 and 8 we only present result for mixtures with the alcohol molar fraction of 0.2. Also note that O-O distance-energy distributions are compared only for the two extremes of the temperature.

With decreasing temperature, two regions (at around 2.8 Å and -5.0 kcal/mol; and between 4.0 and 5.0 Å and +1.0 kcal/mol) became more populated when considering water-water pairs (Figs. 7a and 7b). A similar behavior was found in pure liquid water, and also in ethanol-water mixtures[26]. For water-2-propanol pairs (Fig. 7c and 7d.) less pronounced, but still noticeable effects occur at the same oxygen-oxygen distances.

The pair interaction energy between two 2-propanol molecules (Figs. 7e and 7f) does not show such sensitivity to the changing temperature as water-water pairs. The only notable difference is that the pair interaction energy among water and 2-propanol molecules becomes more negative on decreasing temperature, as it can be seen in Figure 7f at around -6.0 kcal/mol.

Finally, energy minima can be seen at about -3.0 kcal/mol in Figs. 7 that may be set for H-bond definition. In pure liquid water, this is also an accepted value for H-bond definition[25].

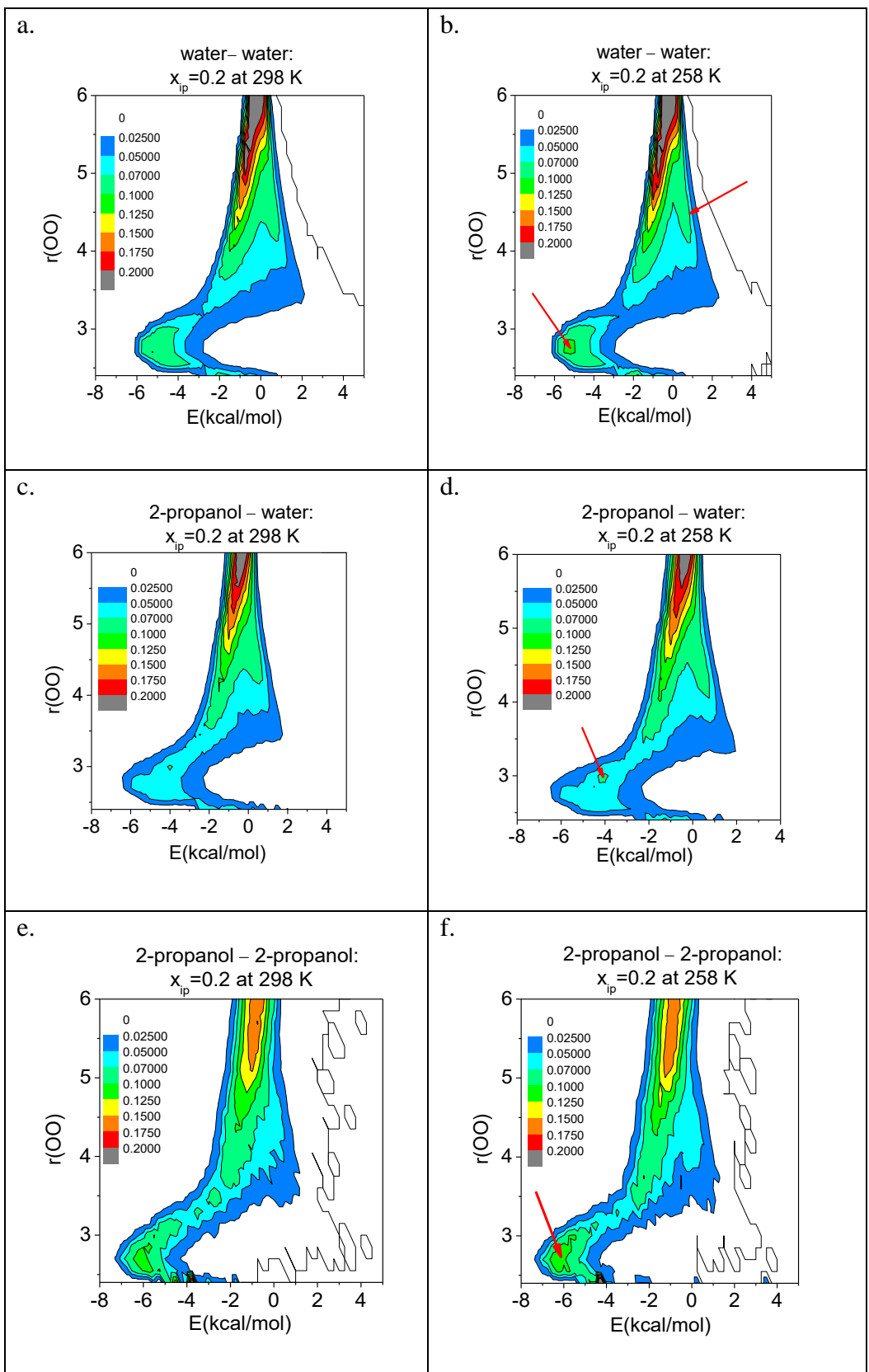

**Figure 7**. Distance-pair energy distributions for water-water, 2-propanol-water and 2-propanol-2-propanol pairs in the mixture with 20 mol % of 2-propanol at 298 K (left panel) and 258 K (right panel). The positions where significant changes may be detected are denoted by red arrows.

In order to explore more clearly the connections and relative configurations of molecules to each other, we calculated spatial distributions (local densities) of the neighboring water and 2-propanol molecules in the first ($r_{oo} < 3.5$ Å) and in the second hydration shells ($3.5$ Å $< r_{oo} < 6$ Å) around central water and 2-propanol molecules. (O atoms are in the origin, x axes defined by the bisectors of the HOH or COH angles, and the xy plane is defined by HOH or COH plane such a way that the z axis is in the positive direction.) As seen in Fig. 8a the local order of water molecules in the first shell is clearly a tetrahedral one (density cutoff=1.15). If we plot the energy distribution function on this surface (first shell), an attractive interaction between -3.0 and -6.0 kcal/mol is found.

On the other hand, the first shell around O atoms of 2-propanol molecules appears in the H-bond donor and acceptor directions, but the structure of second shell is not much ordered (Fig. 8b).

Concerning Figs. 8c and 8d, the typical tetrahedral spatial distribution of neighbors (in the 1$^{st}$ and 2$^{nd}$ shells) around water molecules is clearly preserved also at lower temperature (at 258 K). In Figs 7a and 7b it is demonstrated that the most significant change in this region can be found around 4.0-5.0 Å and +1.0 kcal/mol on the distance-energy map. Using Fig. 8, we can determine that such interactions belong to molecules situated around the bisector direction of the HOH angle in the second shell: in that direction, molecules with a weak repulsive interaction with the central one are positioned more orderly at 258 K (designated by blue color).

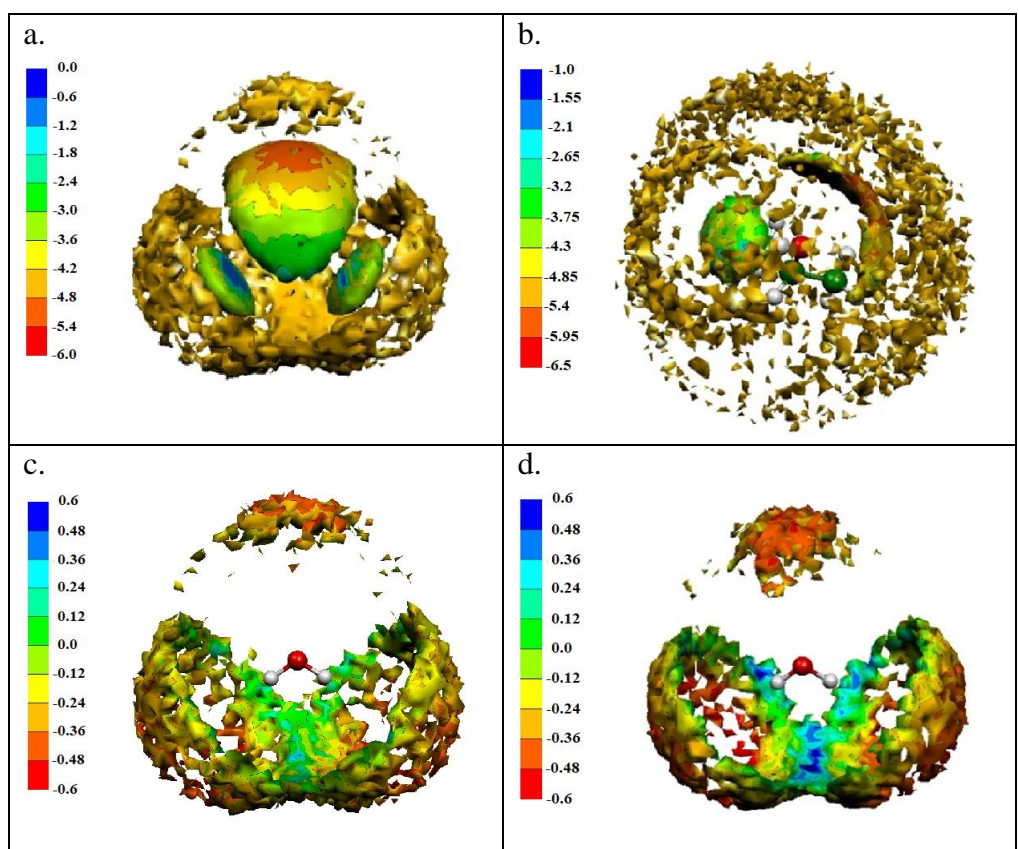

Figure 8 a. First and second shells of the space density distribution for water around water at 298 K. (Water-water energy distribution function is represented by color scale only in the case of first shell). b. First and second shells of the space density distribution for 2-propanol around 2-propanol at 298 K. (2-propanol-2-propanol energy distribution function is represented by color scale only in the case of first shell) c. Second shell of the space density distribution (density cutoff=0.95) for water around water at 298 K (water-water energy distribution function is represented by color coding) d. Second shell of the space density distribution (density cutoff=0.95) for water around water at 258 K (water-water energy distribution function is represented by color coding). All results are shown for $x_{ip}$=0.2.

3.3 Dynamic aspects

3.3.1 Diffusion coefficients

Mean squared displacements (MSD) of centers of mass as a function of time are used here to calculate the self-diffusion coefficient by Einstein's method. Calculated self-diffusion coefficients as a function of temperature are presented in Table 3.

|  | $x_{ip}$= 0.05 | | $x_{ip}$=0.1 | | $x_{ip}$=0.2 | |
|---|---|---|---|---|---|---|
| T(K) | isopropanol | water | isopropanol | water | isopropanol | water |
| 320 | 1.0978 | 2.3348 | 0.852 | 1.8616 | 0.7879 | 1.2783 |
| 298 | 0.6292 | 1.3676 | 0.4624 | 0.9588 | 0.3874 | 0.688 |
| 268 | 0.1655 | 0.4037 | 0.1051 | 0.2535 | 0.072 | 0.1375 |
| 263 | 0.1263 | 0.3094 | 0.08 | 0.1774 | 0.056 | 0.1032 |
| 258 | 0.092 | 0.234 | 0.0549 | 0.1258 | 0.0333 | 0.0785 |
| 253 | 0.0641 | 0.1725 | 0.0304 | 0.0804 | 0.0257 | 0.0535 |
| 248 | 0.0475 | 0.1233 | 0.0197 | 0.0562 | 0.0145 | 0.0309 |
| 243 | 0.0296 | 0.0843 | 0.0141 | 0.0358 | 0.0089 | 0.0196 |
| $T_0$(K) | 197 | 202 | 177 | 198 | 190 | 208 |

Table 3 Calculated self-diffusion coefficients of the components as a function of temperature. (Temperatures where measured structure factors are available are shaded by grey.)

The temperature dependence of $D_w$ and $D_{ip}$ over the temperature range 298 K to 258 K can be described by Arrhenius plots, as shown in Figure 9, although a notable deviation from the 'regular' behavior is also apparent.

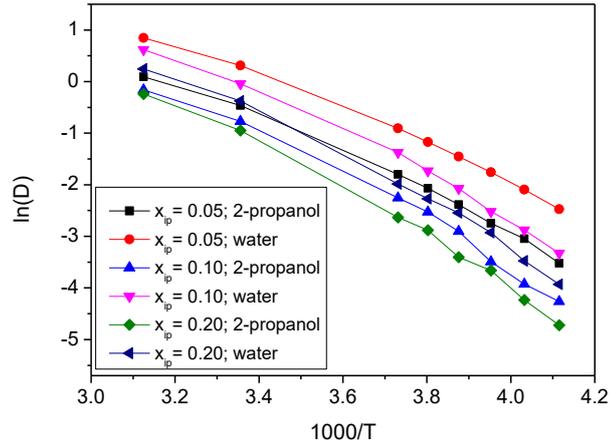

Figure 9 Arrhenius plots related to the self-diffusion coefficients in 2-propanol-water mixtures.

Additionally, we fitted the temperature dependence of the diffusion coefficients of water and 2-propanol molecules by the following functional form, introduced by Rozmanov and Kusalik[59]:

$$D(T) = A \exp\left(\frac{B}{T-T_0}\right) + C$$

where A, B, C, and $T_0$ are optimized parameters. The fitting procedure, due to the quite wide range of D(T), was performed using a modified temperature dependence[59], namely 50D(T)+log(D(T)). The weighting factor (W=50) ensures that the magnitudes of D(T) at higher temperatures and the log(D(T)) values at lower temperatures are numerically comparable.

This D(T) functional form describes the diffusion of a glass forming liquid, with temperature of dynamical arrest $T_0$. Our calculated values for $T_0$ are presented in Table 3. It is valid for every system that $T_0$ is significantly lower than the lowest temperature investigated in this work.

There may be a possibility that water and/or 2-propanol molecules aggregate in hydrogen bonded clusters, which clusters might have different diffusion coefficient themselves. To reveal if such dynamical heterogeneities can be found in terms of the motion of water or 2-propanol molecules, we calculated the $P(r^2)$ probability distribution[27] for different times and temperatures. The highest isopropanol content case ($x_{ip}$=0.2) at the lowest investigated temperature (at 258 K) was chosen to demonstrate our findings (Fig. 10). The shape of these distributions for 2-propanol and water molecules is very similar to each other: all the curves possess a well-defined maximum, a long tail and an expected value of $r^2$ at a certain time. This behavior ($r^2*\exp(-ar^2)$) is the direct consequence of the diffusion law. It means that, at least in terms of this type of motion, no dynamical heterogeneities can be detected in our liquid mixtures. It is worth noting that a very similar behavior was found also in water-ethanol mixtures[26].

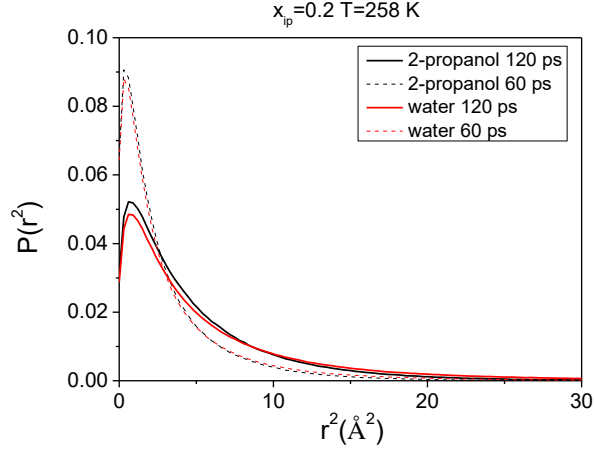

Figure 10 P(r²) probability distribution

### 3.3.2 H-bond dynamics and surviving probability

We have studied the surviving probability (lifetime of H-bond) as calculated according to the following function[54, 60-62]:

$$c_n = \frac{\langle \delta h_n^I(t) \delta h_n^I(0) \rangle}{\langle \delta h_n^I(0) \delta h_n^I(0) \rangle}$$

where

$$\delta h_n^I(t) = h_n^I(t) - \langle h_n^I(t) \rangle$$

The function $h_n^I(t)$ has been defined in the following way:

$$h_n^I(t) = 1$$

if a 2-propanol or water molecule that was in the HB state *n* at time *t*=0 is in the same HB state at time *t*, irrespective of whether or not its HB state has changed in the meantime, and 0 otherwise.

An estimate for the lifetime from this correlation function can be obtained by the following formula[61,62]:

$$\tau_n^I = \int c_n^I \, dt$$

Results for liquid water at 298 and 238 K, using the TIP4P/2005 model, are shown as reference, in Fig 11 (panel 'a').

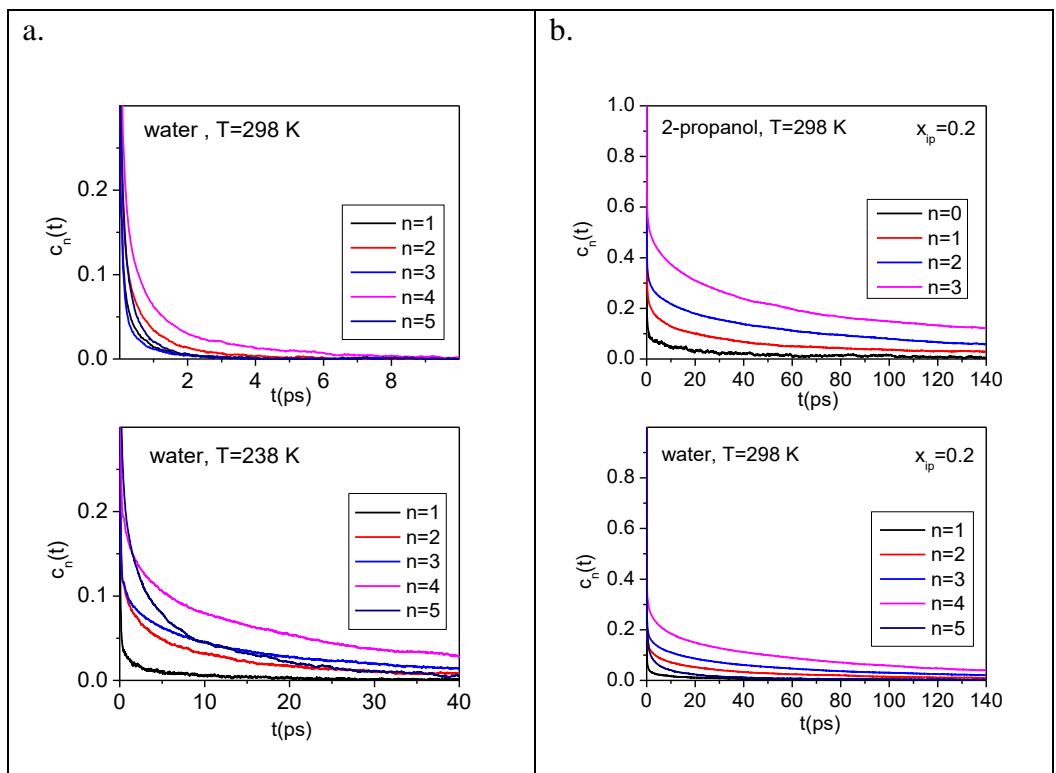

Figure 11 Surviving times (H-bond lifetimes) of the different H-bonding states (according to the number of H-bonded neighbors which is coded by different colors) in pure water, as a function of temperature (panel 'a'); and in the mixture with isopropanol molar ratio of 0.2 at room temperature (panel 'b').

The calculated lifetimes ('survival times') for different H-bonding states in liquid water are presented in Table 4. We can conclude that the n=4 state has significantly longer lifetime than the other states at each investigated temperature.

| T (K) | n=1 | n=2 | n=3 | n=4 | n=5 |
|---|---|---|---|---|---|
| 298 | 0.08 | 0.14 | 0.05 | 0.27 | 0.10 |
| 268 | 0.11 | 0.37 | 0.32 | 0.54 | 0.33 |
| 253 | 0.14 | 0.44 | 0.41 | 0.86 | 0.44 |
| 243 | 0.17 | 0.74 | 0.92 | 1.53 | 0.75 |
| 233 | 0.23 | 1.12 | 1.59 | 2.86 | 1.62 |

Table 4 Calculated lifetimes (ps), as a function of temperature, of water molecules with given numbers of H-bonds in TIP4P/2005 pure water.

The correlation functions $c_n^i(t)$ obtained for water and 2-propanol at room temperature are also shown in Fig. 11 (panel 'b'), and calculated lifetimes are given for the mixture with $x_{ip}=0.2$ in Table 5. It appears that the relaxation process occurs over multiple time scales, as it can be seen in the inset of Fig. 11. All the correlation functions have an initial fast decay over a timescale of about 0.2 ps, followed by a long time exponential decay. Kumar et al.[54] obtained very similar conclusions for liquid water using different H-bond definitions. For each concentration and at each temperature the n=3 and n=4 H-bonded states have the longest lifetimes for 2-propanol and water molecules, respectively.

|  | 2-propanol | | | | water | | | | |
|---|---|---|---|---|---|---|---|---|---|
| T(K) | n=0 | n=1 | n=2 | n=3 | n=1 | n=2 | n=3 | n=4 | n=5 |
| 298 | 1.10 | 1.45 | 1.51 | 4.05 | 0.31 | 0.50 | 0.83 | 1.49 | 0.31 |
| 268 | 2.55 | 5.50 | 10.84 | 19.92 | 0.74 | 3.03 | 4.04 | 7.60 | 1.07 |
| 263 | 2.65 | 6.88 | 12.73 | 23.09 | 0.86 | 3.37 | 4.97 | 10.11 | 1.27 |
| 258 | 2.86 | 8.69 | 16.71 | 29.65 | 0.94 | 4.27 | 7.41 | 13.32 | 1.96 |

Table 5 Calculated lifetimes (ps), as a function of temperature, of 2-propanol (left hand part) and water (right hand part) molecules with given numbers of H-bonds in the mixture with $x_{ip}=0.2$. Note how much longer lived are H-bonds in the mixture than they are in pure water (cf. Table 4).

The temperature dependence of calculated lifetimes for different H-bonded states of water and 2-propanol molecules are presented in Fig. 12, again, for the mixture with $x_{ip}=0.2$. The temperature dependence of these quantities can be reasonably well described by an Arrhenius activation process over the temperature range between 298 and 258 K. Values of the activation energy for all investigated systems are presented in Table 6. The largest activation barrier was found for the n=3 H-bonding state for water, and for the n=2 state for 2-propanol, except for the case of $x_{ip}=0.05$ for 2-propanol.

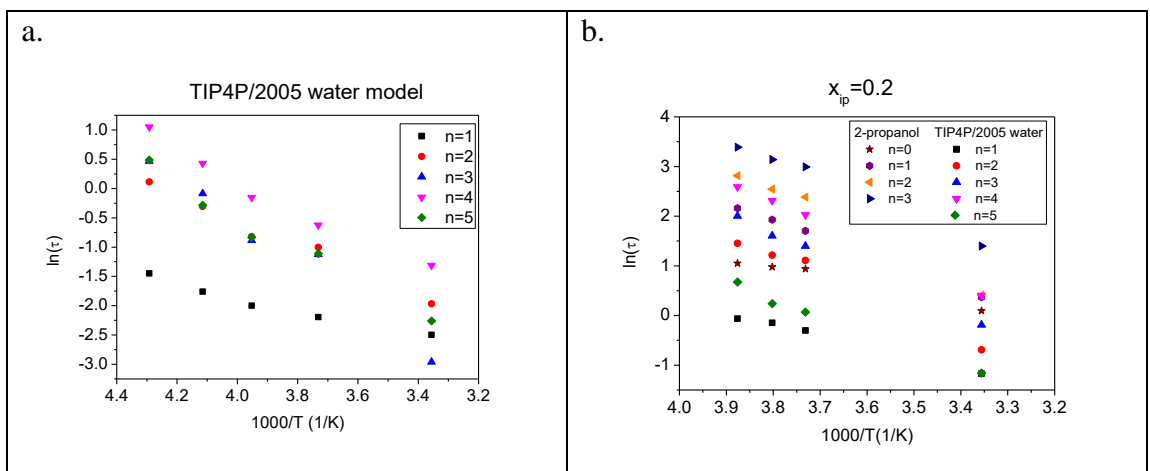

Figure 12. Temperature dependence of the lifetimes of various H-bonding configurations in the mixture with 20 mol% isopropanol (shown as an Arrhenius-plot). (Data for lifetimes can be found in Table 5.)

|  | 2-propanol | | | | water | | | | |
|---|---|---|---|---|---|---|---|---|---|
| $x_{ip}$ | n=0 | n=1 | n=2 | n=3 | n=1 | n=2 | n=3 | n=4 | n=5 |
| 0 |  |  |  |  | 8.98 | 17.88 | 29.10 | 20.62 | 23.03 |
| 0.05 | 1.21 | 40.24 | 19.62 | 31.18 | 6.82 | 13.30 | 22.95 | 17.46 | 19.29 |
| 0.1 | 3.82 | 27.27 | 31.93 | 28.27 | 16.21 | 25.69 | 39.57 | 32.26 | 25.77 |
| 0.2 | 15.88 | 28.85 | 39.24 | 32.26 | 18.21 | 35.09 | 34.50 | 35.17 | 28.02 |

Table 6 Calculated activation barriers (in kJ/mol) in different H-bonded states of isopropanol and water molecules in the investigated systems.

## 4. CONCLUSIONS

Molecular dynamics simulations for 2-propanol-water mixtures, as a function of temperature (between freezing and room temperature) and composition ($x_{ip}$= 0, 0.5, 0.1 and 0.2) have been conducted for temperatures reported in the only available experimental structure study[15]. We have found that:

(1) Comparison with the measured X-ray structure factors revealed that out of the water potentials tested, the TIP4/2005 one[44] provided nearly quantitative agreement with experiment. Therefore for more detailed analyses, particle configurations for this water model have been collected.
(2) Similarly to methanol-water[25] and ethanol-water[26] mixtures, the number of cyclic H-bonded clusters increases on lowering the temperature. The outstanding importance of 6-membered rings observed in methanol-water mixtures is shared here by 5-membered cycles, similarly to ethanol-water mixtures.
(3) Concerning the size of hydrogen-bonded assemblies, not only the mixture as a whole, but also, the water-subsystem is percolating at each temperature and composition studied. On the other hand, only short, isolated chain-like assemblies (consisting of less than 10 molecules) can be found for 2-propanol.
(4) 2-propanol–water H-bond energy distributions (Fig. 6) show a double maximum between -7 and -4 kcal/mol. This feature could not be observed in ethanol-water mixtures.
(5) H-bonding lifetimes in the mixtures tend to be significantly longer than they in pure water.
(6) 'Perfect' H-bonding configurations, i.e. 3 and 4 hydrogen bonds per isopropanol and water molecules, respectively, tend to be the longest lived in the mixtures (and in pure water, too), at each temperature considered here.

Supporting Information

Lennard-Jones parameters and partial charges for the atom types of 2-propanol used in the MD simulations (Table S1); applied temperatures, box lengths, together with the corresponding number densities and bulk densities (Table S2); comparison of the measured (from X-ray diffraction) and calculated (from MD simulations) total scattering structure factors as a function of temperature for the mixture with 5 and 10 mol % isopropanol (Figure S1-S2); heavy-atom related partial radial distribution functions as a function of temperature for the mixture with 5, 10, 20 mol % 2-propanol (Figure S3-S5); H-bond related partial radial distribution functions as a function of temperature for the mixture with 5, 10, 20 mol % 2-propanol (Figure S6-S8); distance-pair energy distributions for water-water, 2-propanol-2-propanol and 2-propanol-water pairs in the mixture with 5 mol % of 2-propanol at 298 K (left panel) and 268 K (right panel) (Figure S9).

Acknowledgement

The authors are grateful to the National Research, Development and Innovation Office (NRDIO (NKFIH), Hungary) for financial support via grants Nos. SNN 116198, KH 130425, 124885 and FK 128656. Sz. Pothoczki acknowledges that this project was supported by the János Bolyai Research Scholarship of the Hungarian Academy of Sciences.